\documentclass[12pt,a4paper,epsf,draft]{article}
\usepackage{rotating}
\usepackage{graphicx}
\usepackage{tabularx}
\usepackage{axodraw}
\def\lsim{\ \rlap{\raise 3pt \hbox{$<$}}{\lower 3pt \hbox{$\sim$}}\ }
\def\gsim{\ \rlap{\raise 3pt \hbox{$>$}}{\lower 3pt \hbox{$\sim$}}\ }

\def\ncr{\langle r_V^2 \rangle}
\def\acr{\langle r_A^2 \rangle}
\def\ncrt{\langle r_V^2(\nu_\tau)  \rangle}
\def\acrt{\langle r_A^2(\nu_\tau) \rangle}
\def\vacrt{\langle r_{V,A}^2(\nu_\tau) \rangle}
\def\ncrm{\langle r_V^2(\nu_\mu) \rangle}
\def\acrm{\langle r_A^2(\nu_\mu) \rangle}

\def\swsq{{\sin^2\theta_W}}

\begin{document}

\title{
  \begin{flushright}
\normalsize
hep-ph/0210137 \\ 
UdeA-PE-02/004\\  
\medskip
 \today \\
  \end{flushright}
\bigskip
Bounds on  the tau and muon neutrino vector and axial vector charge radius
}  

 \author{Martin Hirsch$^{a}$, 
  Enrico Nardi$^{b,c}$ 
and Diego Restrepo$^{c}$ \\ \\ 
\small ${}^a$Instituto de F{\'\i}sica Corpuscular, CSIC - Universitat 
     de Val{\`e}ncia, \\  
\small Edificio Institutos de Paterna, Apartado de Correos 22085\\ 
\small     E-46071--Val{\`e}ncia, Spain \\
\small ${}^b$ INFN-Laboratori Nazionali di Frascati, \\ 
\small C.P. 13, 00044 Frascati, Italy \\
\small ${}^c$ Departamento de F{\'\i}sica,  Universidad de Antioquia \\
\small A.A. {\it 1226} \ Medell{\'\i}n,  \ Colombia 
}

\date{}


\maketitle

\begin{abstract}  
 \noindent
 A Majorana neutrino is characterized by just one flavor diagonal
 electromagnetic form factor: the anapole moment, that in the static limit
 corresponds to the axial vector charge radius $\left< r^2_A\right>$.
 Experimental information on this quantity is scarce, especially in the case
 of the tau neutrino.  We present a comprehensive analysis of the available
 data on the single photon production process $e^+ e^- \to \nu \bar \nu \gamma$ off
 $Z$-resonance, and we discuss the constraints that these measurements can set
 on $\left< r^2_A\right>$ for the $\tau$ neutrino.  We also derive limits for the
 Dirac case, when the presence of a vector charge radius $\left< r^2_V\right>$
 is allowed.  Finally, we comment on additional experimental data on $\nu_\mu$
 scattering from the NuTeV, E734, CCFR and CHARM-II collaborations, and
 estimate the limits implied for $\left< r^2_A\right>$ and $\left<
   r^2_V\right>$ for the muon neutrino.
\end{abstract}

 \newpage


\section{Introduction}
\label{introduction} 

Experimental evidences for neutrino oscillations
 \cite{Toshito:2001dk,Fukuda:2000np,Fukuda:2001nj,Ahmad:2002jz} imply that
neutrinos are the first elementary particles whose properties cannot be fully
described within the Standard Model (SM).  This hints to the
possibility that other properties of these intriguing particles might
substantially deviate from the predictions of the SM, and is presently
motivating vigorous efforts, both on the theoretical and experimental sides,
to understand more in depth the detailed properties of neutrinos and of their
interactions.  In particular, electromagnetic properties of the neutrinos can play
important roles in a wide variety of domains such as cosmology
\cite{Dolgov:2002wy} and astrophysics \cite{Mohapatra:rq,Raffelt:wa}, and
can also provide a viable explanation to the observed depletion of the
electron neutrino flux from the Sun 
\cite{Akhmedov:2002ti,Pulido:2001bd,Miranda:2001hv,Akhmedov:2000fj,Pulido:1999xp,Guzzo:1998sb}.

The electromagnetic interaction of Dirac neutrinos is described
in terms of four form factors. The matrix element of the electromagnetic 
current between an initial neutrino state $\nu_i$ with momentum $p_i$ and 
a final state $\nu_j$ with momentum $p_j$ reads \cite{Nieves:1981zt,Shrock:1982sc}
\begin{eqnarray}
\label{JmuDirac}
\langle \nu^D_j(p_j)\,|\, J_\mu^{\rm EM} \,|\, \nu^D_i(p_i) \rangle &=& i\bar{u}_j \Gamma_\mu^D(q^2) u_i \\ 
 \Gamma_\mu^D(q^2) = (q^2 \gamma_\mu-q_\mu q\!\!\!/) && \hspace{-1truecm}
[V^D(q^2) - A^D(q^2) \gamma_5]  +
i \sigma_{\mu\nu}q^\nu [M^D(q^2)+E^D(q^2)\gamma_5] 
\nonumber
\end{eqnarray}
where $q=p_j-p_i$, and the $(ij)$ indexes denoting the relevant elements of
the form factor matrices have been left implicit.  In the $i=j$ diagonal case,
$M^D$ and $E^D$ are called the magnetic and electric form factors, that in the
limit $q^2=0$ define respectively the neutrino magnetic moment $\mu=M^D(0)$ and
the (CP violating) electric dipole moment $\epsilon=E^D(0)$.  The reduced Dirac form
factor $V^D(q^2)$ and the neutrino anapole form factor $A^D(q^2)$ do not couple
the neutrinos to on-shell photons.  For $i=j$ and in the $q^2 =0$ limit they are
related to the vector and axial vector charge radii $\left< r^2_V \right>$ and
$\left< r^2_A \right>$ through\footnote{The vector charge radius is defined
  as the second moment of the spatial charge distributions $\left< r^2_V
  \right>=\int r^2 \rho_V(r) d\vec r$ where $\rho_V(r)$ is the Fourier transform of the
  full Dirac form factor $q^2 V^D(q^2)$.  The axial vector charge radius can
  be defined in a completely similar way.}
\begin{equation}
\label{radii}
\left< r^2_V \right> = -6\, V^D(0);  \hspace{2truecm} 
\left< r^2_A \right> = -6\, A^D(0).
\end{equation}
In the following, even when $q^2\neq 0$  we will keep referring to the 
reduced Dirac form factor and to the anapole form factor as the 
vector and axial vector charge radius.
A long standing controversy about the possibility of consistently defining
gauge invariant, physical, and process independent vector and axial vector
charge radii \cite{Degrassi:ip} has been recently settled
\cite{Bernabeu:2000hf,Bernabeu:2002nw,Cabral-Rosetti:2002qx,Bernabeu:2002pd}.
The controversy was related to the general problem of defining improved
one-loop Born amplitudes in $SU(2)\times U(1)$ for four fermion processes, like for
example $e^+e^-\to f\bar f$.  If one tries to take into account one-loop vertex
corrections by defining improved effective couplings, one finds that gauge
invariance cannot be preserved unless, together with other one-loop
contributions, $W$ box diagrams are also added to the amplitude.  However, box
diagrams connect initial state fermions to the final states, and thus depend
on the specific process.  Due to the absence of a neutrino-photon coupling at
the tree-level, the problem is even more acute when trying to define the
charge radius as a physical, process independent property, intrinsic to
neutrinos.  In \cite{Bernabeu:2000hf} it was realized that for neutrino
scattering off right handed polarized fermions, the $W$ box diagrams are
absent to begin with, and thus no ambiguity arises. This suggested a way to
derive a unique decomposition of loop contributions that separately respects 
gauge invariance, and from which a process independent charge radius could be
defined as an intrinsic property of the neutrino. Furthermore, in 
\cite{Bernabeu:2002nw,Bernabeu:2002pd} it was
argued that the so-defined charge radius is a physical observable, namely its
value could be extracted, at least in principle, from experiments.

For Majorana neutrinos, in the non-diagonal case ($\nu^M_j\neq\nu^M_i$) and in the
limit of CP invariance the electromagnetic interaction is described by just
two form factors \cite{Nieves:1981zt}.  If the initial and final Majorana
neutrinos involved in the process have the same CP parity, only
$E_{ji}^M(q^2)$ and $A_{ji}^M(q^2)$ are non vanishing, while if the CP parity
is opposite, the electromagnetic interaction is described by $M_{ji}^M(q^2)$
and $V_{ji}^M(q^2)$.  Finally, in the diagonal Majorana case $\nu^M_j=\nu^M_i$ the
only surviving form factor is the anapole moment $A^M(q^2)$.  As discussed
in \cite{Kayser:1982br}, this last result can be inferred from the requirement
that the two identical fermions final state in $\gamma \to \nu^M \bar{\nu}^M$ be
antisymmetric, and therefore it holds regardless of the assumption of CP
invariance.

In the SM the neutrino electromagnetic form factors have extremely small
values \cite{Dubovik:1996gx}.  Due to the left-handed nature of the weak
interactions, the numerical value of the vector and axial vector charge radius
coincide, and for the different $\nu_e$, $\nu_\mu $ and $\nu_\tau $ flavors they fall
within the range $\left< r^2_{V,A} \right> \approx (1 - 4)\times 10^{-33} {\rm cm^2}$
\cite{Bernabeu:2000hf}.\footnote{These values are obtained in the $q^2=0$
  limit, and decrease  with increasing energies with a logarithmic behavior.
}  However, since neutrinos do show properties that are
not accounted for by the SM, it could well be that also their electromagnetic
interactions deviate substantially from the SM expectations.


In general, the strongest limits on the neutrino electromagnetic form factors
come from astrophysical and cosmological considerations.  For example the
neutrino magnetic moments can be constrained from consideration of stellar
energy losses through plasma photon decay $\gamma \to \nu\bar \nu $ \cite{Raffelt:gv},
from the non-observation of anomalous energy loss in the Supernova 1987A
neutrino burst as would have resulted from the rapid emission of superweakly
interacting right handed neutrinos \cite{Raffelt:gv}, and from Big Bang
nucleosynthesis arguments.  In this last case, the agreement between the
measurements of primordial Helium abundance and the standard nucleosynthesis
calculations imply that for example spin flipping Dirac magnetic moment
interactions should be weak enough not to populate right handed neutrinos
degrees of freedom at the time when the neutron-to-proton ratio freezes out
\cite{Dolgov:2002wy}.

Since the charge radii do not couple neutrinos to on-shell photons, the
corresponding interactions are not relevant for stellar evolution arguments.
However, in the Dirac case, right handed neutrinos can still be produced
through e.g. $e^+e^-\to \nu_R\bar \nu_R$, and therefore constraints from the
Supernova 1987A as well as from nucleosynthesis  do apply.  They yield
respectively $|\left< r^2 \right>| \lsim 2\times 10^{-33}{\rm cm}^2$
\cite{Grifols:1989vi} and $|\left< r^2 \right>| \lsim 7\times 10^{-33}{\rm cm}^2$
\cite{Grifols:1986ed}.\footnote{In the SM with right handed
  neutrinos the $\nu_R$ cannot be produced through the charge radius couplings,
  since the vector and axial vector contributions exactly
  cancel. Therefore, the quoted limits implicitly assume that, because of new
  physics contributions, one of the two form factors dominates and no
  cancellations occur.}

However, if neutrinos are Majorana particles, they don't have light
right-handed partners, and the previous constraints do not apply.  In this
case, in particular for the $\tau$ neutrino, an anapole moment corresponding to
an interaction even stronger than electroweak could be allowed.  In the early
Universe such an interaction could keep $\nu_\tau$ in thermal equilibrium long
enough to experience a substantial reheating from $e^+e^- \to \nu_\tau\bar \nu_\tau$
annihilation. We have investigated to what extent this reheating could affect
the Universe expansion rate and change the predictions for primordial Helium
abundance.  As we will discuss in section 2, we have found that even an
interaction one order of magnitude stronger than electroweak would hardly
affect Helium abundance at an observable level.

We conclude that constraints on the Majorana neutrino axial charge radius can
be obtained only from terrestrial experiments.  The present laboratory limits
for the electron neutrino are $-5.5\times 10^{-32} \leq \langle r^2_A (\nu_e) \rangle \leq 9.8\times
10^{-32} {\rm cm}^2$ \cite{Allen:qe}.\footnote{ These limits are twice the
  values published in \cite{Allen:qe} since we are using a convention for
  $\langle r^2_{V,A}\rangle$ that differs for a factor of 2.}  Of course in the Dirac case
these limits apply to the sum $\langle r^2_V \rangle + \langle r^2_A\rangle$ as well.  Limits for the
muon neutrino have been derived from $\nu_\mu$ scattering experiments
\cite{Vilain:1994hm,Ahrens:fp}.  They are about one order of magnitude
stronger than for the electron neutrinos, and will be discussed in section 4.
Due to the fact that intense $\nu_\tau$ beams are not available in laboratories, to
date no direct limits on $\langle r^2_A(\nu_\tau) \rangle $ have been reported by experimental
collaborations.  However, under the assumption that a significant fraction of
the neutrinos from the sun converts into $\nu_\tau$, by using the SNO and
Super-Kamiokande observations the limit $|\langle r^2_A (\nu_\tau) \rangle | \lsim 2\times 10^{-31}
{\rm cm}^2 $ has been derived \cite{Joshipura:2001ee}.  A limit on the $\nu_\tau$
vector charge radius (Dirac case) was derived by analyzing TRISTAN data on the
single photon production process $e^+ e^- \to \nu \bar \nu \gamma$
\cite{Tanimoto:2000am}. The same data can be used to constrain also the
anapole moment for a Majorana $\nu_\tau$, and therefore we have included TRISTAN
measurements in our set of constraints.

In the next section we will briefly analyze the possibility of deriving
constraints on the Majorana neutrino axial charge radius from nucleosynthesis.
In section 3 we will study the bounds on the tau neutrino charge radius
implied by the TRISTAN and LEP experimental results.  In section 4 we will
discuss the constraints on the muon neutrino charge radius from the NuTeV,
CHARM-II, CCFR and the BNL E734 experiments.  They result in the following
90\% c.l.  limits:
\begin{eqnarray}
-8.2 \times 10^{-32}\hskip1mm  {\rm cm}^2 \hskip1mm 
\leq &\left< r^2_A(\nu_\tau) \right>&  \leq
\hskip1mm 9.9 \times 10^{-32} \hskip1mm {\rm cm}^2,  \label{limittau} \\
-5.2 \times 10^{-33}\hskip1mm  {\rm cm}^2 \hskip1mm 
\leq &\left< r^2_A(\nu_\mu) \right>&  \leq
\hskip1mm 6.8 \times 10^{-33} \hskip1mm {\rm cm}^2. \label{limitmu}
\end{eqnarray}
For $\langle r^2_A (\nu_e)\rangle $ we could not find new experimental results that would
imply better constraints than the existing ones \cite{Allen:qe}. We just
mention that the Bugey nuclear reactor data from the detector module closest
to the neutrino source (15 meters) \cite{Declais:1994su} should imply
independent limits of the same order of magnitude than the existing ones.
%
%
 \section{Nucleosynthesis}
 \label{sec:nucleo}
 
 In this section we study the possible impact on the primordial Helium
 abundance $Y$, of an axial charge radius large enough to keep a Majorana
 $\nu_\tau$ in thermal contact with the plasma down to temperatures $T < 1\,$MeV.
 In this case the neutrinos would get reheated by $e^+ e^-$ annihilation, and
 this would affect the Universe expansion rate. To give an example, if one
 neutrino species is maintained in thermal equilibrium until $e^+ e^-$
 annihilation is completed ($T\ll m_e$) this would affect the expansion as $\Delta
 \nu=1-(4/11)^{4/3}\simeq 0.74$ additional neutrinos.
 
 The amount of Helium produced in the early Universe is determined by the
 value of the neutron to proton ratio $n/p$ at the time when the $ne^+ \leftrightarrow p\bar
 \nu$ and $n\nu \leftrightarrow pe^-$ electroweak reactions freeze out.  This occurs
 approximately at a temperature $T_{fo}\approx 0.7\,$MeV \cite{Dicus:bz,Kolb:vq}.
 Apart for the effect of neutron decay, virtually all the surviving neutrons
 end up in $^4He$ nuclei.  Assuming no anomalous contributions to the electron
 neutrino reactions, the freeze out temperature can only be affected by
 changes in the universe expansion rate, that is controlled by the number of
 relativistic degrees of freedom and by their temperature.  If tau neutrinos
 have only standard interactions, at the time of the freeze out they are
 completely decoupled from the thermal plasma. However, an anomalous
 contribution to the process $e^+e^-\leftrightarrow \nu_\tau \bar \nu_\tau$ would allow the $\nu_\tau$ to
 share part of the entropy released in $e^+e^-$ annihilation. The maximum
 effect is achieved assuming that the new interaction is able to keep the
 $\nu_\tau$ thermalized down to $T_{fo}$.  The required strength of the new
 interaction can be estimated by equating the rate for an anomalously fast
 $e^+e^-\leftrightarrow \nu_\tau \bar \nu_\tau$ process $\Gamma_{\nu_\tau}=\left<\sigma v\right> n_e$ to the universe
 expansion rate $\Gamma_{U}= \left(8 \pi \rho/3 m^2_P\right)^{1/2}$.  In the previous
 formulas $\left<\sigma v\right>$ is the thermally averaged cross section times
 relative velocity, $\,n_e\approx 0.365\, T^3$ is the number density of electrons,
 $\rho\approx 1.66\, g_*^{1/2}\, (T^2/m_P) $ is the Universe energy density with $g_*\approx
 10.75$ the number of relativistic degrees of freedom, and $m_P$ is the Plank
 mass.  The thermally averaged cross section can be written as $\left<\sigma
   v\right>\simeq \kappa\, G^2_{\nu_\tau} T^2 $ where $G_{\nu_\tau}\approx (2\pi^2 \alpha/3) \left< r^2_A
 \right>$  parametrizes the strength of the interaction and 
is assumed to be sensibly larger than the Fermi constant $G_F$,
and $\kappa\approx 0.2$ has been
 introduced to allow direct comparison with the SM rate $\left<\sigma
   v\right>^{SM}\simeq 0.2\, G^2_F\, T^2\,$\cite{Dicus:bz}.  By setting
 $\Gamma_{\nu_\tau}=\Gamma_{U}$ at $T=T_{fo}$, we obtain $G_{\nu_\tau}\approx 13 \times 10^{-5}\,$GeV$^{-2}$.
 Therefore, to keep the $\nu_\tau$ thermalized until the ratio $n/p$ freezes out,
 an interaction about ten times stronger than electroweak is needed.
 
 However, even in the presence of such a large interaction, Helium abundance
 would only be mildly affected.  This is because at $T\approx 0.7\,$MeV $e^+e^-$
 annihilation is still not very efficient, and the photon temperature is only
 slightly above the temperature of thermally decoupled neutrinos:
 $(T_\gamma-T_\nu)/T_\gamma\approx1.5\% $ \cite{Dicus:bz}.  This induces a change in the
 primordial Helium abundance $\Delta Y\approx + 0.04\, (\Delta T_{\nu_\tau}/T_\nu)$ which is below
 one part in one thousand.  This effect could possibly be at the level of the
 present theoretical precision \cite{Lopez:1998vk}; however, it is far below
 the present observational accuracy, for which the errors are of the order of
 one percent \cite{Dolgov:2002cv}.

%
%
\section{Limits on $\nu_{\tau}$ vector and axial vector charge radii}
\label{nutau}

Limits on $\ncr$ and $\acr$ for $\nu_{\tau}$ can be set using experimental data on
single photon production through the process $e^+ e^- \to \bar \nu \nu \gamma$.  In the
following we will analyze the data from TRISTAN and the off-resonance data
from LEP. These data have been collected over a large energy range, from 58
GeV up to 207 GeV.  Given that form factors run with the energy, we will
present separate results for the data collected below $Z$ resonance (TRISTAN),
between $Z$ resonance and the threshold for $W^+W^-$ production (LEP-1.5), and
finally for the data above $W^+W^-$ production (LEP-2).  Due to the much
larger statistics collected at high energy, a combined fit of all the data 
does not give any sizable improvement with respect to the LEP-2 limits, that
therefore represent our stronger bounds.

The SM cross section for the process $e^+e^- \to \nu{\bar \nu} \gamma$ is
given by \cite{Gaemers:fe}
\begin{equation}\label{ddsigSM}
\frac{d\sigma_{\nu\nu\gamma}}{dx\,dy} = 
\frac{2\alpha/\pi }{x(1-y^2)}\left[\left(1-\frac{x}{2}\right)^2+\frac{x^2
 y^2}{4} \right]\> \Big\{N_\nu\, \sigma_s(s',g_V,g_A) +  \sigma_{st}(s') + 
\sigma_t(s')\Big\}
\end{equation}
where $\sigma_s$ corresponds to the lowest order $s$ channel $Z$ boson exchange
with $N_\nu=3$ the number of neutrinos that couple to the $Z$ boson.  For later
convenience in $\sigma_s$ we have explicitly shown the dependence on the electron
couplings $g_V=-1/2+2 \sin^2\theta_W$ and $g_A=-1/2$, where $\theta_W$ is the weak
mixing angle.  The additional two terms $\sigma_{st}$ and $\sigma_t$ in (\ref{ddsigSM})
correspond respectively to $Z$-$W$ interference and to $t$ channel $W$ boson
exchange in $\nu_e$ production.  The kinematic variables are the scaled photon
momentum $x=E_{\gamma}/E_{\rm beam}$ with $E_{\rm beam} = \sqrt{s}/2$, the reduced
center of mass energy $s'=s(1-x)$, and the cosine of the angle between the
photon momentum and the incident beam direction $y=\cos\theta_{\gamma}$.  The
expressions for the lowest order cross sections appearing in (\ref{ddsigSM})
read
\begin{eqnarray}
\sigma_s(s) &=& \frac{s\,G_F^2 }{6\pi}\> \frac{\frac{1}{2}\, (g_V^2+g_A^2)\, M^4_Z}
{\left(M^2_Z-s\right)^2+ M_Z^2\Gamma_Z^2}\,, \label{sigs} \\ [5pt]
\sigma_{st}(s) &=& \frac{s\,G_F^2 }{6\pi}\>  \frac{(g_V+g_A)\,(M^2_Z-s)\,M^2_Z}
{\left(M^2_Z-s\right)^2+ M_Z^2\Gamma_Z^2}\,, \label{sigst} \\ [5pt]
\sigma_{t}(s) &=& \frac{s\,G_F^2 }{6\pi}\,,    \label{sigt}
\end{eqnarray}
where $G_F$ is the Fermi constant, $\alpha$ the fine structure constant, $M_Z$ and
$\Gamma_Z$ the mass and width of the $Z$ boson.  Few comments are in order. Eq.
(\ref{ddsigSM}) was first derived in \cite{Gaemers:fe}. It holds at relatively
low energies where $W$ exchange in the $t$ channel can be legitimately
approximated as a contact interaction.  This amounts to neglect the momentum
transfer in the $W$ propagator, and to drop the $W$-$\gamma$ interaction, so that
the photons are emitted only from the electron lines.  While this
approximation is sufficiently good at TRISTAN energies, to analyze the LEP
data collected above $Z$ resonance some improvements have to be introduced.
We will use an improved approximation where finite distance effects are taken
into account in the $W$ propagator, however we will still work in the limit of
vanishing $W$-$\gamma$ interactions. While strictly speaking the amplitude with
photon attached only to the electron legs is not gauge invariant, the
necessary contribution for completing the gauge invariant amplitude is of
higher order in a leading log approximation \cite{Bardin:2001vt}, and for our
scopes can be safely neglected.  Finite distance $W$ exchange effects can be
taken into account in the previous expressions through the replacement
\begin{eqnarray}
\label{finitedist}
\sigma_{st}(s) & \to &  \sigma_{st}(s) \cdot F_{st} \left(\frac{s}{M^2_W}\right) \label{distancest}\\  [5pt] 
\sigma_{t}(s) & \to &  \sigma_{t}(s) \cdot F_{t} \left(\frac{s}{M^2_W}\right)   \label{distancet}
\end{eqnarray}
where $M_W$ is the $W$ boson mass, and 
\begin{eqnarray}
F_{st}(z) &=& \frac{3}{z^3} \Big[(1+z)^2\log(1+z) -
z\,\Big(1+\frac{3}{2}\,z\Big) \Big]\,, \label{fst}  \\  
F_{t}(z)  &=&  \frac{3}{z^3} \Big[-2\,(1+z)\,\log(1+z) + z\,(2+z) \Big]\,.   \label{ft}
\end{eqnarray}
The contact interaction approximation is recovered in the limit $z\to 0$ for
which $F_{st,t}(z) \to 1$.

An anomalous interaction due to non-vanishing $\nu_{\tau}$ axial and axial vector
charge radii can be directly included in (\ref{ddsigSM}) by redefining the $Z$
boson exchange term in the following way:
\begin{equation}\label{newPhys}
N_\nu\, \sigma_s(s',g_V,g_A) \to (N_\nu-1)\,\sigma_s(s',g_V,g_A)  +\sigma_s(s', g_V^*(s'),g_A).  
\end{equation}
where
\begin{eqnarray}
 g_V^*(s') &=& g_V - \left[1-\frac{s'}{M^2_Z}\right]\, \delta\,, \label{Deftildev}\\ 
\delta &=& \frac{\sqrt{2}\pi\alpha}{3 G_F}\left[\ncr +\acr\right]\,. \label{DefDelta}
\end{eqnarray}
The substitution $g_V\to g_V^*$ in (\ref{newPhys}) takes into account the new
photon exchange diagram for production of left-handed $\nu_\tau$.  In the Dirac
case, $s$-channel production of right handed $\nu_\tau$ through photon exchange
must also be taken into account.  This yields a new contribution that adds
incoherently to the cross section, and that can be included by adding inside
the brackets in (\ref{ddsigSM}) the term
\begin{eqnarray} 
\sigma_R\,(s') &=&  \frac{s'\,G_F^2 }{6\pi}\,  (\delta')^2\,, \\ \label{nuR}
\delta' &=& \frac{\sqrt{2}\pi\alpha}{3 G_F}\left[\ncr -\acr\right].\label{DefDeltaP}
\end{eqnarray}
In the SM $\ncr = \acr$ and therefore there is no production of
$\nu_R$ through these couplings.  For a Majorana neutrino $\delta'=0$ and $\ncr = 0$,
and thus the limits on anomalous contributions to the process $e^+e^- \to \nu
{\bar \nu} \gamma$ translate into direct constraints on the axial charge radius $\acrt$.
Note that including anomalous contributions just for the $\nu_\tau$ is justified by
the fact that for $\nu_e$ and $\nu_{\mu}$ the existing limits are generally stronger than
what can be derived from the process under consideration.

\subsection{Limits from TRISTAN}

The three TRISTAN experiments AMY \cite{Sugimoto:1995bf}, TOPAZ \cite{Abe:fq}
and VENUS \cite{Hosoda:1994bd} have searched for single photon production in
$e^+e^-$ annihilation at a c.m. energy of approximately $\sqrt{s}=58\,$GeV.
Anomalous contributions to the cross section for $e^+e^- \to \nu{\bar \nu} \gamma$ would
have been signaled by an excess of events in their measurements.  Limits on
the tau neutrino charge radius from the TRISTAN data have already been derived
in \cite{Tanimoto:2000am}.  In the present analysis, we include also the
neutrino axial charge radius, and we give an alternative statistical treatment
based on a $\chi^2$-analysis and on the measured cross sections, rather than on
the number of events observed combined with Poisson statistics as given in
\cite{Tanimoto:2000am}. This puts the TRISTAN constraints on a comparable
statistical basis with the LEP results discussed in the next section.

TRISTAN data are collected in table \ref{TristanData}.  The number of single
photon observed, including the SM backgrounds, was respectively 6
for AMY, 5 for TOPAZ, and 8 for VENUS. The numbers listed in the $N_{\rm obs}$
column in table \ref{TristanData} are the background subtracted events, that
correspond to the measured cross sections $\sigma^{\rm mes}$ given in the fourth
column.  We have found that our expressions for the cross section
(\ref{ddsigSM})-(\ref{ft}) tend to overestimate the Monte Carlo results quoted
by the three collaborations. This might be due to additional specific
experimental cuts besides the ones quoted in the last two columns in table
\ref{TristanData} . In any case, the disagreements with the Monte Carlo
results remain well below the experimental errors, and therefore we simply
consider it as an additional theoretical uncertainty that we add in
quadrature. In constructing the $\chi^2$ function, we use conservatively as
experimental errors the upper figures of the three measurements. This is
justified by the fact that the $\gamma-Z$ interference term arising from new
physics is always sub dominant with respect to the square of the anomalous
photon exchange diagram, and therefore new physics contributions would always
increase the cross section.

%
\begin{table}[h] 
\caption{
Summary of the TRISTAN data: The center of mass energy and luminosity are
given in the second and third column. The background subtracted experimental
cross sections and the Monte Carlo expectations quoted by the three
collaborations are given respectively in column four and five (in femtobarns),
while the number of observed events after background subtraction is listed in
column six. $\epsilon$ is the efficiency of the cuts in percent units.  The last two
columns collect the  kinematic cuts, with $x=E_\gamma/E_{\rm beam}$, $x_T=x
\sin \theta_\gamma$ with $\theta_\gamma$ the angle between the photon momentum and the beam
direction, and $y=\cos \theta_\gamma$.
}    
\label{TristanData} 
\setlength{\tabcolsep}{1mm}
\vspace{3mm}
\begin{tabular}{|c|c|c|c|c|c|c|c|c|c}
\hline 
\vbox{\vskip15pt}
  & $\sqrt{s}$ {\footnotesize [GeV]} &{\footnotesize $\cal L\,$[pb$^{-1}$]} &
  $\sigma^{\rm mes}$ [fb] & $\sigma^{\rm MC}$ [fb] &$N_{\rm obs}$
& $\epsilon\,(\%)$ & $E_\gamma/E_{\rm beam}$  & $|y|$  \\  [3pt]
\hline \vbox{\vskip15pt}
 &  & 55 &  $29^{+25}_{-18}$   &34 &  & $44$  &  $x$$\geq$$0.175$ &
   \\  [3pt]
 &  & 91 & {\footnotesize\rm for}  & 34 & & $64$  &  $x$$\geq$$0.175$ &
   \\  [-5pt]
 {\footnotesize AMY \cite{Sugimoto:1995bf}}& 57.8 & &&&
  $4.2^{+3.7}_{-2.6}$ $^a$ &&    &  $\leq$0.7    \\  [-5pt]
&  & 56 & {\footnotesize ($x$$\geq$$0.125$ }  & 49 &  & $58$  &  $x$$\geq$$0.125$ &
   \\  [3pt] 
&  & 99 &  {\footnotesize $|y|\leq$0.7) }   & 49 & & $57$  &  $x$$\geq$$0.125$ &
    \\  [3pt]
\hline\hline  \vbox{\vskip15pt}
{\footnotesize TOPAZ \cite{Abe:fq} } 
&58&213& $37^{+58}_{-19}$ &54 & $2.2^{+3.4}_{-1.1}$ $^b$ & $27.3$ &
                                                   $ x\geq 0.14$ & $\leq$0.8    \\  [3pt]
&  &   & &   &    &  &   $ x_T\geq 0.12$  &   \\  [3pt]
\hline\hline \vbox{\vskip15pt}
 {\footnotesize VENUS \cite{Hosoda:1994bd}} 
     & 58 & 164.1 &$42.0^{+45.3}_{-30.2}$&36.4&$3.9^{+4.2}_{-2.8}$ $^c$  & 57 &
  $x_T$$\geq$0.13 &   $\leq$0.64  \\  [3pt]
\hline 
\end{tabular}
\phantom{aaa} \\
\phantom{aaa} \\
$^a$\footnotesize{
  AMY observes 6 events in the 4 runs listed above (respectively 0, 2, 2, 2)
  with an estimated background of $1.7\pm0.3$ events.  The quoted value for
  $N_{\rm obs}$ has been derived from their
  background subtracted cross-section.}\\
$^b$\footnotesize{
  TOPAZ observes 5 events, and expects $2.5^{+1.5}_{-0.6}$ from background.
  $N_{\rm obs}$ has been derived from their
  background subtracted cross section. }\\
$^c$\footnotesize{ 
  VENUS observes 8 events and expects $4.1^{+2.4}_{-1.7}$ from background.
  They quote $3.9^{+4.2}_{-2.8}$ background subtracted $\bar \nu \nu \gamma$ events,
  which correspond to the cross section given in the fourth column.}\\
\end{table}
%

For a Majorana $\nu_\tau$ ($\delta'=0$ and $\ncr = 0$) the TRISTAN data imply the
following 90 \% c.l.  limits:
\begin{equation}
-3.7 \times 10^{-31}\hskip2mm  {\rm cm}^2 \hskip2mm \leq \acrt \leq
\hskip2mm 3.1 \times 10^{-31} \hskip2mm {\rm cm}^2.  
\label{TristanMaj}
\end{equation}
For the Dirac case, the associated production of right-handed states through
$\sigma_R$ in (\ref{nuR}) allows us to constrain independently the vector and axial
vector charge radius.  The 90 \% c.l.  are:
\begin{equation}  \label{vacrtTRIS} 
-2.1 \times 10^{-31}\hskip2mm  {\rm cm}^2 \hskip2mm \leq \vacrt \leq
\hskip2mm 1.8 \times 10^{-31} \hskip2mm {\rm cm}^2.               
\end{equation}
As we have already mentioned, strictly speaking the constraints just derived
cannot be directly compared with the LEP constraints analyzed below, since the
two experiments are proving neutrino form factors at different energy scales.
Of course, since our limits are meaningful only to the extent that they are
interpreted as constraints on physics beyond the SM, it is not possible to
make a sound guess of the form of the scaling of the form factors with the
energy, which is determined by the details of the underlying new physics.
However, if we assume a logarithmic reduction of the form factors with
increasing energy as is the case in the SM, than we would expect a moderate
reduction of about $\approx 0.65$ when scaling from TRISTAN to LEP-1.5 energies, and
an additional reduction of about $\approx 0.75$ from LEP-1.5 up to LEP-2
measurements at 200 GeV.

\subsection{Limits from LEP}

\nocite{Buskulic:1996hw,Adam:1996am,Alexander:1996kp,Ackerstaff:1997ze,Abbiendi:1998yu}
Limits on $\ncr$ and $\acr$ can be derived from the observation of single
photon production at LEP in a completely similar way. We stress that contrary
to magnetic moment interactions that get enhanced at low energies with respect
to electroweak interactions, the interaction corresponding to a charge radius
scale with energy roughly in the same way than the electroweak interactions,
and therefore searches for possible effects at high energy are not in
disadvantage with respect to low energy experiments.  It is for this reason
that LEP data above the $Z$ resonance are able to set the best constraints on
the vector and axial vector charge radius for the $\tau$ neutrino.

All LEP experiments have published high statistics data for the process
$e^+e^-\to \nu\bar\nu \gamma$ for c.m. energies close to the $Z$-pole; however, due to
the dominance of resonant $Z$ boson exchange, these data are not useful to
constrain anomalous neutrino couplings to $s$-channel off-shell photons.
Therefore, in the following we will analyze LEP data on single photon
production collected above $Z$ resonance, in the energy range 130 GeV -- 207
GeV. We divide the data into two sets: LEP-1.5 data collected below $W^+W^-$
production threshold are collected in table \ref{Lep1Data}, while LEP-2 data,
collected above $W^+W^-$ threshold and spanning the energy range 161 -- 207
GeV are collected in table \ref{Lep2Data}.

\subsubsection{LEP-1.5}

The ALEPH \cite{Buskulic:1996hw}, DELPHI \cite{Adam:1996am} and OPAL
\cite{Alexander:1996kp,Ackerstaff:1997ze,Abbiendi:1998yu} collaborations have
published data for single photon production at c.m. energies of 130 GeV and
136 GeV.  During the fall 1995 runs ALEPH \cite{Buskulic:1996hw} and DELPHI
\cite{Adam:1996am} accumulated about 6 pb$^{-1}$ of data for each
experiment, observing respectively 40 and 37 events.  In the same runs OPAL
\cite{Alexander:1996kp,Ackerstaff:1997ze} collected a little less than 5
pb$^{-1}$ observing 53 events.  In addition, OPAL published data also for the
1997 runs (at the same energies) \cite{Abbiendi:1998yu} collecting an
integrated luminosity of 5.7 pb$^{-1}$ and observing 60 events.

ALEPH reports two values for the cross sections at 130 GeV and 136 GeV, each
based on $2.9$ pb$^{-1}$ of statistics. They also quote the results of a Monte
Carlo calculation of the SM cross section, that is in good agreement with the
experimental numbers (and with our estimates).  DELPHI combined together the
statistics of both the 130 GeV and 136 GeV runs, however they present separate
results for two different detector components: the High density Projection
Chamber (HPC) covering large polar angles, and the Forward ElectroMagnetic
Calorimeter (FEMC) covering small polar angles. Since DELPHI does not quote
any Monte Carlo result we assign a bona fide 5\% theoretical error for our
cross section estimates.  OPAL published two sets of data. The data recorded
in the 1995 runs \cite{Alexander:1996kp} were reanalyzed in
\cite{Ackerstaff:1997ze}, and correspond to $2.30$ pb$^{-1}$ collected at 130
GeV, and to $2.59$ pb$^{-1}$ collected at 136 GeV.  In the 1997
runs\cite{Abbiendi:1998yu} $2.35$ pb$^{-1}$ were collected at 130 GeV, and
$3.37$ pb$^{-1}$ at 136 GeV.  With a total integrated luminosity of about 28
pb$^{-1}$ LEP-1.5 implies the following 90 \% c.l.  limits:
\begin{equation}
-5.9 \times 10^{-31}\hskip2mm  {\rm cm}^2 \hskip2mm \leq \acrt \leq
\hskip2mm 6.6 \times 10^{-31} \hskip2mm {\rm cm}^2 
\label{LEP-1.5Maj}
\end{equation}
for the axial vector charge radius of a Majorana $\nu_\tau $, and 
\begin{equation}   \label{vacrtLEP-1.5}   
 -3.5 \times 10^{-31}\hskip2mm  {\rm cm}^2 \hskip2mm \leq \vacrt \leq
 \hskip2mm 3.7 \times 10^{-31} \hskip2mm {\rm cm}^2             
 \end{equation}
 for the Dirac case.  Let us note that, in spite of the much larger
 statistics, the limits from LEP-1.5 (\ref{LEP-1.5Maj}) and
 (\ref{vacrtLEP-1.5}) are roughly a factor of two worse than the limits from
 TRISTAN in (\ref{TristanMaj}) and (\ref{vacrtTRIS}). The main reason for this
 is that at LEP-1.5 energies initial state radiation tends to bring the
 effective c.m. energy of the collision $s'$ close to the $Z$ resonance, thus
 enhancing $Z$ exchange with respect to the new photon exchange diagram.
%
%
\begin{table}[h] 
\caption{
  Summary of the ALEPH, DELPHI and OPAL data collected below $W^+W^-$
  production threshold.  ALEPH \protect{\cite{Buskulic:1996hw}} and OPAL
   \protect{\cite{Ackerstaff:1997ze,Abbiendi:1998yu}} present separate results for two
  different energies, while DELPHI  \protect{\cite{Adam:1996am}} combines together the
  data collected at 130 and 136 GeV. DELPHI presents separate data for two
  different detector components: the High density Projection Chamber (HPC)
  covering large polar angles, and the Forward Electromagnetic Calorimeter
  (FEMC) covering the forward regions.  The kinematic cuts applied are given
  in columns eight and nine.  Wherever a double error is listed, the first is
  statistical and the second is systematic.
}  
\label{Lep1Data} 
\setlength{\tabcolsep}{1mm}
\vspace{3mm}
\begin{tabular}{|c|c|c|c|c|c|c|c|c|c}
\hline \vbox{\vskip15pt}  LEP-1.5
&$\sqrt{s}${\footnotesize [GeV]}&{\footnotesize $\cal L\,$[pb$^{-1}$]}&$\sigma^{\rm mes}$ [pb]& 
$\sigma^{\rm MC}$ [pb] &N$_{\rm obs}$& $\epsilon\,(\%)$ &$E_\gamma$ {\footnotesize [GeV]}&$|y|$\\  [3pt]
\hline \vbox{\vskip15pt}
{\footnotesize ALEPH } & 130 & 2.9 &9.6$\pm$2.0$\pm$0.3  & 10.7$\pm$0.2 & 23 & $85$  &&   \\  [-5pt]
& && && && $ \geq 10$  & $ \leq 0.95$      \\  [-5pt]
{\footnotesize  \protect{\cite{Buskulic:1996hw}}} 
&136 &2.9 & 7.2$\pm$1.7$\pm$0.2&\ 9.1$\pm$0.2&17&$85$&& \\  [3pt]
\hline\hline \vbox{\vskip15pt}
{\footnotesize DELPHI  } & && && && &              \\  [0pt]
{\footnotesize HPC  \protect{\cite{Adam:1996am}}}
&$\langle133\rangle$ &5.83&7.9$\pm$1.9$\pm$0.7&-&20&$53^*$&$\geq2$\ \ & $\leq0.70$\\ [2pt]
{\footnotesize FEMC\protect{\cite{Adam:1996am}}}
&$\langle133\rangle$ &5.83&6.0$\pm$1.9$\pm$0.6&-&17&$43^*$&$\geq10$&$0.83$-$0.98$\\ [3pt]
\hline\hline \vbox{\vskip15pt}
{\footnotesize OPAL}
&130&2.30&10.0$\pm$2.3$\pm$0.4&13.48$\pm$0.22$^{\dagger}$&19&$81.6$&  $x_T$$>$$\,0.05$&$\leq0.82$\\ [-5pt]
& && && &&    $\!\!\!${\footnotesize  or} \qquad\qquad & \\  [-5pt]
{\footnotesize\cite{Ackerstaff:1997ze}}   
&136 &2.59 &16.3$\pm$2.8$\pm$0.7&11.30$\pm$0.20$^{\dagger}$&34&$79.7$&  $x_T$$>$$\,0.1\,$ & $\leq0.966$    \\ [2pt]
\hline \vbox{\vskip15pt}
 & 130 &2.35 & 11.6$\pm$2.5$\pm$0.4   & 14.26$\pm$0.06$^{\dagger}$ & 21 & $77.0 $ && \\  [-5pt]
{\footnotesize\protect{\cite{Abbiendi:1998yu}}} & && && && $x_T$$>$$\,0.05$ & $\leq0.966$  \\  [-5pt]
&136 &3.37 & 14.9$\pm$2.4$\pm$0.5&11.95$\pm$0.07$^{\dagger}$& 39 & $77.5 $  & & \\  [2pt]
\hline 
\end{tabular}
\phantom{aaa} \\
\phantom{aaa} \\
$^*$\footnotesize{
Estimated   from the inferred experimental  cross sections and measured numbers of events.}\\
$^{\dagger}$\footnotesize{
Calculated from the expected number of events as predicted by  the KORALZ event generator.}\\
\end{table}
%
\subsubsection{LEP-2}

Above the threshold for $W^+W^-$ production the four LEP experiments collected
altogether about 1.6 nb$^{-1}$ of data. The corresponding 24 data-points are
collected in table (\ref{Lep2Data}).  
ALEPH \cite{Barate:1997ue,Barate:1998ci,Heister:2002ut} published data for ten
different c.m. energies, ranging from 161 GeV up to 209 GeV.  Data collected
between 203.0 GeV and 205.5 GeV were combined together, they appear in
the table as the 205 GeV entry, and the same was done for the data collected
between 205.5 GeV and 209.0 GeV that are quoted as the 207 GeV entry.  DELPHI
\cite{Abreu:2000vk} published data collected at 183 GeV and 189 GeV, and gives
separate results for the three major electromagnetic calorimeters, the HPC,
the FEMC and  the Small angle TIle Calorimeter (STIC) that covers the very
forward regions, between $2^\circ - 10^\circ$ and $170^\circ - 178^\circ$.  In
three papers \cite{Acciarri:1997dq,Acciarri:1998hb,Acciarri:1999kp} L3
reported the results obtained at 161 GeV, 172 GeV, 183 GeV and 189 GeV.  While
for most data points the agreement between our SM computation of the
cross-sections and the Monte Carlo results is at the level of 5 \% or better,
we find that the L3 Monte Carlo results are up to 20\% larger than our
numbers, and this disagreement is encountered for all the four L3 data points.
While we have not been able to track the reasons of this discrepancy, we have
verified that the effects on our final results is negligible.  OPAL published
data for four different c.m. energies
\cite{Ackerstaff:1997ze,Abbiendi:1998yu,Acciarri:1999kp}.  For the data
presented in \cite{Ackerstaff:1997ze,Abbiendi:1998yu} we have estimated the
Monte Carlo cross sections from the published numbers of events expected as
predicted by the KORALZ event generator. The results agree well with our
estimates.

The  90 \% c.l. limits implied by   LEP-2 data read 
\begin{equation}
-8.2 \times 10^{-32}\hskip2mm  {\rm cm}^2 \hskip2mm \leq \acrt \leq
\hskip2mm 9.9 \times 10^{-32} \hskip2mm {\rm cm}^2 
\label{LEP-2Maj}
\end{equation}
%
%
%
\begin{table}[p] 
\caption{
  Summary of the ALEPH, DELPHI, L3 and OPAL experimental data, collected above
  $W^+W^-$ production threshold.  The notation is the same than in
  (\ref{Lep1Data}).  Wherever a double error is listed, the first is
  statistical and the second is systematic.  
}
\label{Lep2Data} 
\setlength{\tabcolsep}{1mm}
\vspace{2mm}
\begin{tabular}{|c|c|c|c|c|c|c|c|c|c}
\hline \vbox{\vskip15pt} LEP-2 
  & $\sqrt{s}$ {\footnotesize [GeV]} &{\footnotesize $\cal L\,$[pb$^{-1}$]} &
  $\sigma^{\rm mes}$ [pb] & $\sigma^{\rm MC}$ [pb] &N$_{\rm obs}$
& $\epsilon\,(\%)$ & $E_\gamma$ {\footnotesize [GeV]} & $|y|$  \\  [2pt]
\hline \vbox{\vskip15pt}
{\footnotesize ALEPH }
 & 161 & 11.1 &5.3$\pm$0.8$\pm$0.2  & 5.81$\pm$0.03 & 41 & $70$  &&   \\  [-7pt]
& && && && $ x_T\geq 0.075$  & $ \leq 0.95$      \\  [-7pt]
 {\footnotesize \cite{Barate:1997ue} }
 & 172 & 10.6 &4.7$\pm$0.8$\pm$0.2  & 4.85$\pm$0.04 & 36 & $72$  && \\  [2pt]
\hline \vbox{\vskip15pt}
{\footnotesize\cite{Barate:1998ci}}&183&58.5&4.32$\pm$0.31$\pm$0.13&4.15$\pm$0.03&195&$77$&
                                                   $ x_T\geq 0.075$ & $ \leq 0.95$    \\  [2pt]
\hline  \vbox{\vskip15pt}
 & 189 & 173.6 & 3.43$\pm$0.16$\pm$0.06 & 3.48$\pm$0.05 & 484 & &&   \\  [2pt]
 & 192 &\ 28.9 & 3.47$\pm$0.39$\pm$0.06 & 3.23$\pm$0.05 &\ 81 & &&   \\  [2pt]
 & 196 &\ 79.9 & 3.03$\pm$0.22$\pm$0.06 & 3.26$\pm$0.05 & 197 & &&   \\  [2pt]
 {\footnotesize\cite{Heister:2002ut} }
 & 200 &\ 87.0 & 3.23$\pm$0.21$\pm$0.06 & 3.12$\pm$0.05 & 231 & 81.5& $ x_T\geq 0.075$  & $ \leq 0.95$      \\  [2pt]
 & 202 &\ 44.4 & 2.99$\pm$0.29$\pm$0.05 & 3.07$\pm$0.05 & 110 & &&   \\  [2pt]
 & 205 &\ 79.5 & 2.84$\pm$0.21$\pm$0.05 & 2.93$\pm$0.05 & 182 & &&   \\  [2pt]
 & 207 & 134.3 & 2.67$\pm$0.16$\pm$0.05
 & 2.80$\pm$0.05 & 292 & &&   \\  [2pt]
\hline\hline \vbox{\vskip13pt}
{\footnotesize DELPHI  } 
& && && && &              \\  [-2pt]
{\footnotesize \cite{Abreu:2000vk}}  & && && && &              \\  [-7pt]
&183&\ 50.2&1.85$\pm$0.25$\pm$0.15   & 2.04 &\ 54&$58^\sharp$& & \\ [-7pt]
{\footnotesize HPC } & && && &&  $x\geq0.06$\ \ & $\leq0.70$\\ [-7pt]
                    &189& 154.7&1.80$\pm$0.15$\pm$0.14& 1.97 & 146&$51^\sharp$& & \\ [-5pt]
{\footnotesize  } & && && && &              \\  [-6pt]
&183&\ 49.2&2.33$\pm$0.31$\pm$0.18& 2.08 &\ 65& $54^\sharp$ & $x\geq$0.2 & $\geq$0.85 \\ [-7pt]
{\footnotesize FEMC } & && && && &   \\ [-7pt]           
    &189& 157.7&1.89$\pm$0.16$\pm$0.15& 1.94 &\ 155& $50^\sharp$&$x\leq$0.9 &  $\leq$0.98 \\ [5pt]
&183&\ 51.4&1.27$\pm$0.25$\pm$0.11& 1.50 &\ 32& -- $^\ddagger$ & $x\geq$0.3 & $\geq$0.990 \\ [-7pt]
{\footnotesize STIC } & && && && &   \\ [-7pt]           
    &189& 157.3&1.41$\pm$0.15$\pm$0.13& 1.42 &\ 94& -- $^\ddagger$&$x\leq$0.9 & $\leq$0.998 \\ [2pt]
\hline\hline \vbox{\vskip15pt}
 {\footnotesize L3 }&161&10.7 &6.75$\pm$0.91$\pm$0.18&6.26$\pm$0.12&57&$80.5$&$\geq10$&$\leq0.73$ \\ [-2pt]
&&&&&&&$\!\!${\footnotesize  and}\qquad\qquad & \\  [-2pt]
{\footnotesize\cite{Acciarri:1997dq}}&172&10.2&6.12$\pm$0.89$\pm$0.14&\ 5.61$\pm$0.10&49&$80.7$&
                                      $E_T\geq6$&  $0.80$--$0.97$\\  [2pt]
\hline \vbox{\vskip15pt}
 {\footnotesize  \cite{Acciarri:1998hb}} &183 
& 55.3 &5.36$\pm$0.39$\pm$0.10 & 5.62$\pm$0.10 & 195 & $65.4$  & $\geq 5$ & $\leq0.73$   \\  [-3pt]
&&&&&&&$\!\!${\footnotesize and}\qquad\qquad & \\  [-3pt]
 {\footnotesize  \cite{Acciarri:1999kp}} &189 
& 176.4 &5.25$\pm$0.22$\pm$0.07 & 5.29$\pm$0.06 & 572 & $60.8$  & $E_T\geq5$&  $0.81$--$0.97$\\  [3pt]
\hline\hline \vbox{\vskip15pt}
{\footnotesize OPAL}
&161 & 9.89 & 5.3$\pm$0.8$\pm$0.2 & 6.49$\pm$0.08$^{\dagger}$&40& $75.2$   
                                                 &  $x_T$$>$$\,0.05$&$\leq0.82$\\ [-5pt]
&&&&&&&$\!\!${\footnotesize or}\qquad\qquad & \\  [-5pt]
{\footnotesize\cite{Ackerstaff:1997ze}}
 &172 &10.28 & 5.5$\pm$0.8$\pm$0.2&  5.53 $\pm$0.08$^{\dagger}$&45& $77.9$ & $x_T$$>$$\,0.1\,$ &  $\leq0.966$\\ [3pt]
\hline \vbox{\vskip15pt}
 {\footnotesize\cite{Abbiendi:1998yu}} 
 &183 &54.5 & 4.71$\pm$ 0.34$\pm$0.16 & 4.98$\pm$0.02$^{\dagger}$ & 191 & $74.2 $  & 
  $x_T$$>$$\,0.05$&  $\leq0.966$ \\  [3pt]
\hline \vbox{\vskip15pt}
 {\footnotesize \cite{Abbiendi:2000hh}} &189 &177.3 &
  4.35$\pm$0.17$\pm$0.09&4.66$\pm$0.03 
                  & 643 & $82.1 $  &  $x_T$$>$$\,0.05$&  $\leq0.966$ \\  [3pt]
\hline 
\end{tabular}
\phantom{aaa} \\
$^\ddagger$  \footnotesize
{The STIC Calorimeter efficiency 
varies between 74\% and 27\% over the angular region used in the analysis. }\\
$^\sharp$\footnotesize{
Estimated   from the Monte Carlo  cross sections and the expected numbers of events.}\\
$^{\dagger}$\footnotesize
{Calculated from the expected number of events as predicted by  the KORALZ event generator.}\\
\end{table}
%
%
for the Majorana case, and   
\begin{equation} \label{vacrtLEP-2}  
-5.6 \times 10^{-32}\hskip2mm  {\rm cm}^2 \hskip2mm \leq \vacrt \leq
\hskip2mm 6.2 \times 10^{-32} \hskip2mm {\rm cm}^2               
\end{equation}
for a Dirac $\nu_\tau$.

\begin{figure}
\vskip-10mm
\hskip35mm
\epsfysize=80mm
\epsfxsize=80mm
 \epsfbox{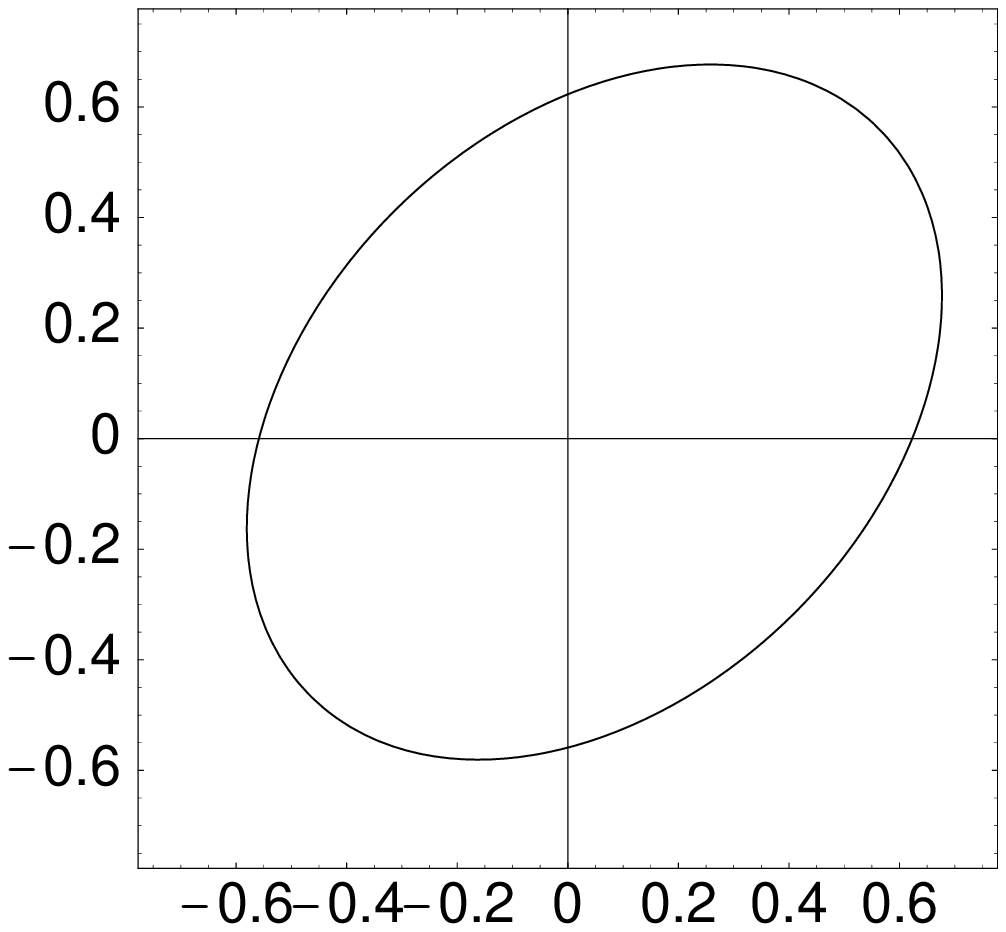}
\vskip-35mm
 \hskip30mm
\begin{rotate}{90}
$\ncrt$ [$10^{-31}$ cm$^2$]
\end{rotate}

\vskip30mm
 \hskip65mm
$\acrt$ [$10^{-31}$ cm$^2$]
\caption[]{Combined limits on $\ncrt$ and $\acrt$ for Dirac tau neutrinos 
derived from LEP-2 data. The plot shows the 
$\chi^2_{\rm min}+2.71$ contour, corresponding to 90 \% c.l.}
\label{Lep2Lim}
\end{figure}

These limits are about a factor of four stronger than the limits derived in
\cite{Joshipura:2001ee} from the SNO and Super-Kamiokande observations and
than the limits obtained in \cite{Tanimoto:2000am} from just the TRISTAN data.
In Fig.~\ref{Lep2Lim} we depict the 90 \% c.l. limits on $\ncrt$ and $\acrt$
for the Dirac case as derived from the LEP-2 data. The picture shows the
absence of any strong correlation between $\ncrt$ and $\acrt$.  We stress that
the possibility of bounding simultaneously the vector and axial vector charge
radii stems from the fact that in $e^+ e^-$ annihilation also the right-handed
neutrinos can be produced, and they couple to the photon through a combination
of $\ncr$ and $\acr$ which is orthogonal to the one that couples the
left-handed neutrinos.  In contrast, neutrino scattering experiments do not
involve the right handed neutrinos, and therefore can only bound the
combination $\ncr + \acr$.

Before concluding this section, we should mention that independent limits
could also be derived from the DONUT experiment, through an analysis similar to
the one presented in \cite{Schwienhorst:2001sj}, and that yielded 
limits on the $\nu_\tau$ magnetic moment.  We have estimated that the constraints
from DONUT would be at least one order of magnitude worse than the limits
obtained from LEP; however, it should be remarked that these limits would be
inferred directly from the absence of anomalous interactions for a neutrino
beam with an identified $\nu_\tau$ component \cite{Kodama:2000mp}.
%
%
%
%

\section{Limits on $\nu_{\mu}$ vector and axial vector charge radius}

The NuTeV collaboration has recently published a value of $\swsq$ 
measured from the ratio of neutral current to charged current 
in deep inelastic $\nu_{\mu}$-nucleon scattering  \cite{Zeller:2001hh}.
Their result reads
\begin{equation}
\sin^2\theta_W^{(\nu)} = 0.2277 \pm 0.0013 \pm 0.0009
\label{NuTeVRes}
\end{equation}
where the first error is statistical and the second error is systematic. 
In order to derive limits on neutrino electromagnetic properties one should
compare the results obtained in neutrino experiments to a value of $\swsq$
determined from experiments that do not involve neutrinos.  Currently, the most
precise value of $\swsq$ from non-neutrino experiments comes from measurements
at the $Z$-pole 
and from direct measurements of the $W$-mass \cite{Hagiwara:pw}. 
In our numerical calculations we will use the value for $\swsq$ 
obtained from a global fit to electroweak measurements without 
neutrino-nucleon scattering data, as reported in 
\cite{Zeller:2001hh,Zeller:2002dx}:
\begin{equation}
\swsq = 0.2227 \pm 0.00037\,.
\label{EWWGsw}
\end{equation}
The effect of a non-vanishing charge radius can be taken into account through
the replacement $g_V \to g_V - \delta$ in the formulas for $\nu_\mu$-nucleon and
$\nu_\mu$-electron scattering \cite{Vogel:iv}, where $\delta$ is given in
(\ref{DefDelta}).  Since there are no right-handed neutrinos involved, there
is no effect proportional to $\delta'$ and therefore only $\delta \propto \ncrm + \acrm$ can
be constrained.  Upper and lower limits can be directly derived by comparing
$\sin^2\theta_W^{(\nu)}$ with the quoted value of $\swsq$ from non-neutrino experiments.
Since the results for neutrino experiments and the measurements at the
$Z$-pole are not consistent at the 1$\sigma$ level, in the following equations
(\ref{LimNuTeV})-(\ref{LimCharm}) we will (conservatively) combine the errors
by adding them linearly.\footnote{Except for the CCFR data, which is 
consistent with the SM precision fits.}
%
\begin{figure}
\vskip0mm
\hskip15mm
\epsfysize=60mm
\epsfxsize=100mm
\epsfbox{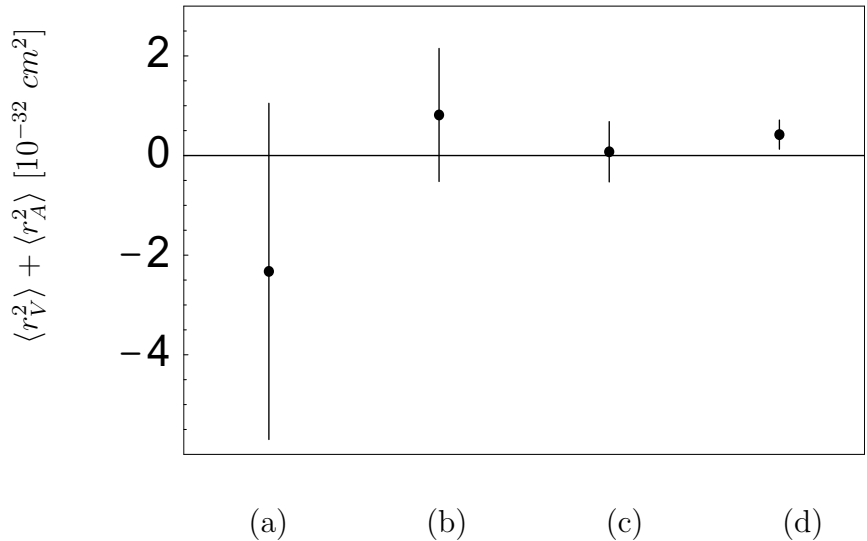}

\vskip-20mm
\hskip5mm
\begin{rotate}{90}
$\ncr+\acr$ [$10^{-32}$ $cm^2$]
\end{rotate}

\vskip20mm
\hskip33mm
(a) \hskip17mm (b) \hskip17mm (c) \hskip17mm (d) 

\caption[]{90 \% c.l. limits on ($\ncr$+$\acr$) for the muon neutrino 
derived from (a) E734 at BNL \cite{Ahrens:fp}, 
(b) CHARM II \cite{Vilain:1994hm}, (c) CCFR experiment 
\cite{McFarland:1997wx} and (d) from the
  NuTeV result \cite{Zeller:2001hh}.}
\label{NuTevLim}
\end{figure}
%
%

From the  NuTeV  result (\ref{NuTeVRes}) we obtain the 90 \% c.l. upper limit:
\begin{equation}
\ncrm + \acrm \leq
\hskip2mm 7.1 \times 10^{-33} \hskip2mm {\rm cm}^2.  
\label{LimNuTeV}
\end{equation}
A reanalysis of the E734 data on $\nu_\mu$-$e$ and $\bar \nu_\mu$-$e$
scattering \cite{Ahrens:fp} yields the  90 \% c.l. limits:
\begin{equation}
-5.7 \times 10^{-32}\hskip2mm  {\rm cm}^2 \hskip2mm \leq \ncrm + \acrm \leq
\hskip2mm 1.1 \times 10^{-32} \hskip2mm {\rm cm}^2.   
\label{LimAhr}
\end{equation}
Note that in ref. \cite{Ahrens:fp} the E734 collaboration is quoting a lower
limit about 3.6 times and an upper limit about 7.5 times tighter than the ones
given in (\ref{LimAhr}). This is because of various reasons: first of all, as
was pointed out in \cite{Allen:1990xn}, in \cite{Ahrens:fp} an inconsistent
value for $G_F$ was used that resulted in bounds stronger by approximately a
factor of $\sqrt{2}$. In addition, the errors were combined quadratically,
which, due to the large negative trend in their data, resulted in a much
stronger upper bound on $\ncrm+\acrm$ than the one quoted here.  Finally, our
value of $\delta$ is defined through the shift $g_V \to g_V - \delta$ of the SM vector
coupling, consistently for example with the notation of \cite{Vogel:iv}, while
the convention used by the E734 Collaboration \cite{Ahrens:fp} as well as by
CHARM II \cite{Vilain:1994hm} define $\delta$ as a shift in $\swsq$. This implies
that our limits are larger for an additional factor of 2 with respect to the
results published by these two collaborations.

From the  CHARM II neutrino-electron scattering data
\cite{Vilain:1994hm} we obtain  at 90 \% c.l.: 
\begin{equation}
-0.52 \times 10^{-32}\hskip2mm  {\rm cm}^2 \hskip2mm \leq \ncrm + \acrm \leq
\hskip2mm 2.2 \times 10^{-32} \hskip2mm {\rm cm}^2.  
\label{LimCharm}
\end{equation}
These limits differ from the numbers published by the CHARM II collaboration
\cite{Vilain:1994hm} not only because of the mentioned factor of 2 in the 
definition of $\delta$, but also  because the present value of $\swsq$ \cite{Hagiwara:pw}
is smaller than the one used in 1995 in the CHARM II analysis.  

From the data published by the CCFR collaboration \cite{McFarland:1997wx} 
one can deduce
\begin{equation}
-0.53 \times 10^{-32}\hskip2mm  {\rm cm}^2 \hskip2mm \leq \ncrm + \acrm \leq
\hskip2mm 0.68 \times 10^{-32} \hskip2mm {\rm cm}^2.  
\label{LimCCFR}
\end{equation}

The four limits discussed above are represented in fig. \ref{NuTevLim},  
that makes apparent the level of precision of the NuTeV result.  
By combining the upper limit from CCFR (\ref{LimCCFR}) and the 
lower limit from CHARM II (\ref{LimCharm}) we finally obtain:
\begin{equation}
-5.2 \times 10^{-33}\hskip2mm  {\rm cm}^2 \hskip2mm \leq \ncrm + \acrm \leq
\hskip2mm 6.8 \times 10^{-33} \hskip2mm {\rm cm}^2.  
\label{LimAll}
\end{equation}
It is well known that the NuTeV result shows a sizable deviation from the SM
predictions \cite{Zeller:2001hh}, and as a consequence it also appears to be
inconsistent (at the 90 \% c.l.) with $\delta = 0$. In fact, strictly speaking
their result 
$\ncrm+\acrm = (4.20\pm 1.64)\times 10^{-33}$ cm$^2$ (1 $\sigma$ error)
could be interpreted as a measurement of $\ncrm+\acrm$. A vanishing value for
$\ncrm+\acrm$ becomes consistent with NuTeV data only at approximately $2.5$
standard deviations.
We should also mention that the fact that the NuTeV central value is very
close to the range for $\ncr$ at $q^2=0$ quoted in the introduction should not
mislead to think that a SM effect has been measured.  In the SM the charge
radius $\left<r^2 (q^2)\right>$ runs from its value at $q^2=0$ approximately
as $\log(|q|^2/M_W^2)$.  In the NuTeV experiment the energy transfer is always
$>20$ GeV \cite{Zeller:2001hh}, and therefore at the typical interaction
energies of this experiment the value of the charge radius is expected to be
smaller than its value in the static limit by at least a factor of ten.

%
\vspace{-3truemm}
\section{Conclusions}
\label{conclusions} 

This work stems from the observation that if neutrinos are Majorana particles
their axial charge radius $\acr\,$, that is the only permitted flavor diagonal
electromagnetic form factor, cannot be constrained through astrophysical or
cosmological observations. In section 2 we have discussed in some
detail how it is not possible to derive useful constraints from
nucleosynthesis and from the measurements of primordial Helium abundance.
We have concluded that in order to constrain $\acr$ we can rely only
on the analysis of the results of terrestrial experiments.

In section 3 we have presented a comprehensive analysis of the available off
$Z$-resonance data for the process $e^+ e^- \to \nu \bar \nu \gamma$.  We have used these
data to derive limits for the axial vector charge radius of the $\tau$ neutrino,
as well as combined limits on the vector and axial vector charge radius in the
case of a Dirac $\nu_\tau$.  These limits are largely dominated by the high
statistics LEP-2 data collected above $W^+W^-$ production threshold.

We have also analyzed the bounds that can be derived for the muon neutrino
from an analysis of neutrino scattering experiments.  We obtained the most
stringent limits by combining the CCFR $\nu_\mu$-nucleon scattering and the CHARM
II $\nu_\mu$-electron scattering results.  No new limits were obtained for the
electron neutrino; however, new experiments dedicated to the detailed study of
electron (anti)neutrino interactions with matter, as for example the MUNU
experiment at the Bugey nuclear reactor \cite{Broggini:sa}, should be able to
improve existing limits by about one order of magnitude.

\vspace{-3truemm}
\section{Acknowledgments}
\label{acknowledgments} 

This work was supported in part by COLCIENCIAS in Colombia and by CSIC in
Spain through a joint program for international scientific cooperation, and in
part by the Spanish grant BFM2002-00345 and by the European Commission RTN
network HPRN-CT-2000-00148.  M. H.  is supported by a Spanish MCyT Ramon y
Cajal contract.


\bigskip



\begin{thebibliography}{99}



\bibitem{Toshito:2001dk}
T.~Toshito  [Super-Kamiokande Collaboration],
arXiv:hep-ex/0105023.

\bibitem{Fukuda:2000np}
S.~Fukuda {\it et al.}  [Super-Kamiokande Collaboration],
Phys.\ Rev.\ Lett.\  {\bf 85} (2000) 3999
[arXiv:hep-ex/0009001].

\bibitem{Fukuda:2001nj}
S.~Fukuda {\it et al.}  [Super-Kamiokande Collaboration],
Phys.\ Rev.\ Lett.\  {\bf 86} (2001) 5651
[arXiv:hep-ex/0103032].

\bibitem{Ahmad:2002jz}
Q.~R.~Ahmad {\it et al.}  [SNO Collaboration],
Phys.\ Rev.\ Lett.\  {\bf 89} (2002) 011301
[arXiv:nucl-ex/0204008].

\bibitem{Dolgov:2002wy}
A.~D.~Dolgov,
arXiv:hep-ph/0202122.

\bibitem{Mohapatra:rq}
R.~N.~Mohapatra and P.~B.~Pal,
World Sci.\ Lect.\ Notes Phys.\  {\bf 60} (1998) 1.

\bibitem{Raffelt:wa}
G.~G.~Raffelt,
{\it  Chicago, USA: Univ. Pr. (1996) 664 p}.

\bibitem{Akhmedov:2002ti}
E.~K.~Akhmedov and J.~Pulido,
Phys.\ Lett.\ B {\bf 529} (2002) 193
[arXiv:hep-ph/0201089].

\bibitem{Pulido:2001bd}
J.~Pulido,
arXiv:hep-ph/0112104.

\bibitem{Miranda:2001hv}
O.~G.~Miranda, C.~Pena-Garay, T.~I.~Rashba, V.~B.~Semikoz and J.~W.~Valle,
Phys.\ Lett.\ B {\bf 521} (2001) 299
[arXiv:hep-ph/0108145]; Nucl.\ Phys.\ B {\bf 595} (2001) 360 [arXiv:hep-ph/0005259].
%
\bibitem{Akhmedov:2000fj}
E.~K.~Akhmedov and J.~Pulido,
Phys.\ Lett.\ B {\bf 485} (2000) 178
[arXiv:hep-ph/0005173].
%
\bibitem{Pulido:1999xp}
J.~Pulido and E.~K.~Akhmedov,
Astropart.\ Phys.\  {\bf 13} (2000) 227
[arXiv:hep-ph/9907399].
%
\bibitem{Guzzo:1998sb}
M.~M.~Guzzo and H.~Nunokawa,
Astropart.\ Phys.\  {\bf 12} (1999) 87
[arXiv:hep-ph/9810408].







\bibitem{Nieves:1981zt}
J.~F.~Nieves,
Phys.\ Rev.\ D {\bf 26} (1982) 3152.




\bibitem{Shrock:1982sc}
R.~E.~Shrock,
Nucl.\ Phys.\ B {\bf 206} (1982) 359.

\bibitem{Degrassi:ip}
See for example: 
G.~Degrassi, A.~Sirlin and W.~J.~Marciano,
Phys.\ Rev.\ D {\bf 39} (1989) 287; 
M.~J.~Musolf and B.~R.~Holstein,
Phys.\ Rev.\ D {\bf 43} (1991) 2956.
A complete list can be found in \cite{Bernabeu:2000hf}.

\bibitem{Bernabeu:2000hf}
J.~Bernabeu, L.~G.~Cabral-Rosetti, J.~Papavassiliou and J.~Vidal,
Phys.\ Rev.\ D {\bf 62} (2000) 113012
[arXiv:hep-ph/0008114].

\bibitem{Bernabeu:2002nw}
J.~Bernabeu, J.~Papavassiliou and J.~Vidal,
Phys.\ Rev.\ Lett.\  {\bf 89}, 101802 (2002), 
[arXiv:hep-ph/0206015].   

\bibitem{Cabral-Rosetti:2002qx}
L.~G.~Cabral-Rosetti, M.~Moreno and A.~Rosado,
arXiv:hep-ph/0206083.

\bibitem{Bernabeu:2002pd}
J.~Bernabeu, J.~Papavassiliou and J.~Vidal,
arXiv:hep-ph/0210055.


\bibitem{Kayser:1982br}
B.~Kayser,
Phys.\ Rev.\ D {\bf 26} (1982) 1662.


\bibitem{Dubovik:1996gx}
V.~M.~Dubovik and V.~E.~Kuznetsov,
Int.\ J.\ Mod.\ Phys.\ A {\bf 13} (1998) 5257
[arXiv:hep-ph/9606258].

\bibitem{Raffelt:gv}
G.~G.~Raffelt,
Phys.\ Rept.\  {\bf 320} (1999) 319.




\bibitem{Grifols:1989vi}
J.~A.~Grifols and E.~Masso,
Phys.\ Rev.\ D {\bf 40} (1989) 3819.

\bibitem{Grifols:1986ed}
J.~A.~Grifols and E.~Masso,
Mod.\ Phys.\ Lett.\ A {\bf 2} (1987) 205.

\bibitem{Allen:qe}
R.~C.~Allen {\it et al.},   
Phys.\ Rev.\ D {\bf 47} (1993) 11. 


\bibitem{Vilain:1994hm}   
P.~Vilain {\it et al.}  [CHARM-II Collaboration],
Phys.\ Lett.\ B {\bf 345} (1995) 115.  

\bibitem{Ahrens:fp}    
L.~A.~Ahrens {\it et al.} [E734 Collaboration], 
Phys.\ Rev.\ D {\bf 41} (1990) 3297.


\bibitem{Joshipura:2001ee}
A.~S.~Joshipura and S.~Mohanty,
arXiv:hep-ph/0108018.

\bibitem{Tanimoto:2000am}
N.~Tanimoto, I.~Nakano and M.~Sakuda,
Phys.\ Lett.\ B {\bf 478} (2000) 1
[arXiv:hep-ph/0002170].


\bibitem{Declais:1994su}
Y.~Declais {\it et al.},
Nucl.\ Phys.\ B {\bf 434} (1995) 503.


\bibitem{Dicus:bz}
D.~A.~Dicus, E.~W.~Kolb, A.~M.~Gleeson, E.~C.~Sudarshan, V.~L.~Teplitz and M.~S.~Turner,
Phys.\ Rev.\ D {\bf 26} (1982) 2694.

\bibitem{Kolb:vq}
E.~W.~Kolb and M.~S.~Turner,
{\it  Redwood City, USA: Addison-Wesley (1990) 547 p. (Frontiers in physics, 69)}.

\bibitem{Lopez:1998vk}
R.~E.~Lopez and M.~S.~Turner,
Phys.\ Rev.\ D {\bf 59} (1999) 103502
[arXiv:astro-ph/9807279].


\bibitem{Dolgov:2002cv}
A.~D.~Dolgov,
Nucl.\ Phys.\ Proc.\ Suppl.\  {\bf 110} (2002) 137
[arXiv:hep-ph/0201107].

\bibitem{Gaemers:fe}
K.~J.~Gaemers, R.~Gastmans and F.~M.~Renard,
Phys.\ Rev.\ D {\bf 19} (1979) 1605.

\bibitem{Bardin:2001vt}
D.~Bardin, S.~Jadach, T.~Riemann and Z.~Was,
Eur.\ Phys.\ J.\ C {\bf 24} (2002) 373
[arXiv:hep-ph/0110371].

\bibitem{Sugimoto:1995bf}
Y.~Sugimoto {\it et al.}  [AMY Collaboration],
Phys.\ Lett.\ B {\bf 369} (1996) 86.

\bibitem{Abe:fq}
T.~Abe {\it et al.}  [TOPAZ Collaboration],
Phys.\ Lett.\ B {\bf 361} (1995) 199.

\bibitem{Hosoda:1994bd}
N.~Hosoda {\it et al.}  [VENUS Collaboration],
Phys.\ Lett.\ B {\bf 331} (1994) 211.


\bibitem{Buskulic:1996hw}
D.~Buskulic {\it et al.}  [ALEPH Collaboration],
Phys.\ Lett.\ B {\bf 384} (1996) 333.

\bibitem{Adam:1996am}
W.~Adam {\it et al.}  [DELPHI Collaboration],
Phys.\ Lett.\ B {\bf 380} (1996) 471.


\bibitem{Alexander:1996kp}
G.~Alexander {\it et al.}  [OPAL Collaboration],
Phys.\ Lett.\ B {\bf 377} (1996) 222.

\bibitem{Ackerstaff:1997ze}
K.~Ackerstaff {\it et al.}  [OPAL Collaboration],
Eur.\ Phys.\ J.\ C {\bf 2} (1998) 607
[arXiv:hep-ex/9801024]. 


\bibitem{Abbiendi:1998yu}
G.~Abbiendi {\it et al.}  [OPAL Collaboration],
Eur.\ Phys.\ J.\ C {\bf 8} (1999) 23
[arXiv:hep-ex/9810021].  

\bibitem{Barate:1997ue}
R.~Barate {\it et al.}  [ALEPH Collaboration],
Phys.\ Lett.\ B {\bf 420} (1998) 127
[arXiv:hep-ex/9710009].

\bibitem{Barate:1998ci}
R.~Barate {\it et al.}  [ALEPH Collaboration],
Phys.\ Lett.\ B {\bf 429} (1998) 201.

\bibitem{Heister:2002ut}
A.~Heister {\it et al.}  [ALEPH Collaboration],
CERN-EP-2002-033.


\bibitem{Abreu:2000vk}
P.~Abreu {\it et al.}  [DELPHI Collaboration],
Eur.\ Phys.\ J.\ C {\bf 17} (2000) 53
[arXiv:hep-ex/0103044].


\bibitem{Acciarri:1997dq}
M.~Acciarri {\it et al.}  [L3 Collaboration],
Phys.\ Lett.\ B {\bf 415} (1997) 299.


\bibitem{Acciarri:1998hb}
M.~Acciarri {\it et al.}  [L3 Collaboration],
Phys.\ Lett.\ B {\bf 444} (1998) 503.

\bibitem{Acciarri:1999kp}
M.~Acciarri {\it et al.}  [L3 Collaboration],
Phys.\ Lett.\ B {\bf 470} (1999) 268
[arXiv:hep-ex/9910009].


\bibitem{Abbiendi:2000hh}
G.~Abbiendi {\it et al.}  [OPAL Collaboration],
Eur.\ Phys.\ J.\ C {\bf 18} (2000) 253
[arXiv:hep-ex/0005002].

\bibitem{Schwienhorst:2001sj}
R.~Schwienhorst {\it et al.}  [DONUT Collaboration],
Phys.\ Lett.\ B {\bf 513} (2001) 23
[arXiv:hep-ex/0102026].


 \bibitem{Kodama:2000mp}
 K.~Kodama {\it et al.}  [DONUT Collaboration],
 Phys.\ Lett.\ B {\bf 504} (2001) 218
 [arXiv:hep-ex/0012035].


\bibitem{Zeller:2001hh}
G.~P.~Zeller {\it et al.}  [NuTeV Collaboration],
Phys.\ Rev.\ Lett.\  {\bf 88} (2002) 091802
[arXiv:hep-ex/0110059].

\bibitem{Hagiwara:pw}
K.~Hagiwara {\it et al.}  [Particle Data Group Collaboration],
Phys.\ Rev.\ D {\bf 66} (2002) 010001.

\bibitem{Zeller:2002dx}
G.~P.~Zeller  [NuTeV Collaboration],
arXiv:hep-ex/0207037.


\bibitem{Vogel:iv}
P.~Vogel and J.~Engel,
Phys.\ Rev.\ D {\bf 39} (1989) 3378.

\bibitem{McFarland:1997wx}
K.~S.~McFarland {\it et al.}  [CCFR Collaboration],
Eur.\ Phys.\ J.\ C {\bf 1} (1998) 509
[arXiv:hep-ex/9701010].

\bibitem{Allen:1990xn}
R.~C.~Allen {\it et al.},  
 Phys.\ Rev.\ D {\bf 43} (1991) R1.


\bibitem{Broggini:sa}
C.~Broggini  [MUNU Collaboration],
Nucl.\ Phys.\ Proc.\ Suppl.\  {\bf 110} (2002) 398.













































\end{thebibliography}
\end{document}